\documentclass[pra,reprint,amsmath,amssymb,aps,floatfix]{revtex4-1}

\usepackage{graphicx}
\usepackage{dcolumn}
\usepackage{bm}
\usepackage{hyperref}
\usepackage[mathlines]{lineno}
\usepackage{todonotes}

\usepackage{amsthm}
\usepackage{tabularx}
\usepackage{float}
\usepackage{tikz}
\usepackage{tikz-cd}

\newcommand{\Bb}{\mathbb{B}}

\newcommand{\Nb}{\mathbb{N}}

\newcommand{\Rb}{\mathbb{R}}

\newcommand{\Zb}{\mathbb{Z}}

\newcommand{\coim}{\mathrm{\coim}}

\newcommand{\wtil}{\widetilde}

\newcommand{\abs}[1]{\lvert #1 \rvert}

\newcommand{\hook}{\hookrightarrow}

\newcommand{\Deck}{\mathrm{Deck}}
\newcommand{\Map}{\mathrm{Map}}
\newcommand{\ev}{\mathrm{ev}}
\newcommand{\RP}{\mathbb{R}\mathrm{P}}
\newcommand{\SO}{\mathrm{SO}}

\newtheorem{theo}{Tplottin ubuntuheorem}[section]
\theoremstyle{plain}
\newtheorem{thm}[theo]{Theorem}
\newtheorem{lem}[theo]{Lemma}
\newtheorem{prop}[theo]{Proposition}
\newtheorem{cor}[theo]{Corollary}

\newtheorem*{thm*}{Theorem}
\newtheorem*{lem*}{Lemma}
\newtheorem*{prop*}{Proposition}
\newtheorem*{cor*}{Corollary}

\theoremstyle{definition}
\newtheorem{defn}[theo]{Definition}
\newtheorem{ex}[theo]{Example}

\newtheorem{rem}[theo]{Remark}

\begin{document}
\title{Charge ambiguity and splitting of monopoles}
\author{Toni Annala}\email{toni.annala@aalto.fi}\affiliation{QCD Labs, QTF Centre of Excellence, Department of Applied Physics, Aalto University, P.O. Box 13500, FI-00076 Aalto, Finland}
\author{Mikko M{\"o}tt{\"o}nen}\affiliation{QCD Labs, QTF Centre of Excellence, Department of Applied Physics, Aalto University, P.O. Box 13500, FI-00076 Aalto, Finland}
\date{\today}

\begin{abstract}
This paper is dedicated to studying various aspects of topological defects, appearing in mean-field theory treatments of physical systems such as ultracold atomic gases and gauge field theories. We start by investigating topological charge ambiguity and addition of topological charges using the mathematical formalism of covering spaces, which clarifies many aspects these phenomena. Subsequently, we classify topological defect configurations consisting of several monopoles and unknotted ring defects in terms of homotopy groups and fundamental group actions on them, thus generalizing the previous classifications of a single monopole and a single unknotted ring defect. Finally, we examine the decay of multiply charged topological monopoles under small perturbations of the physical system, and analyze the conditions under which multiply charged monopoles are inclined to split into several singly charged monopoles.
\end{abstract}


\maketitle

\section{Introduction}

For the past century, the theory of topology has been fruitfully applied \cite{ghrist:2014} in a variety of computational \cite{GraphBook} and physical problems \cite{Nakahara:2003}. For example, it has been used in the study of graph colorings \cite{Babson:2004, Huh:2012}, computational electromagnetics \cite{gross:2004}, and provided ways to classify materials and their properties in terms of the lattice structure \cite{Mermin:1979} or the structure of electronic wave-functions in conductors \cite{Thouless:1983}. Curiously, topological defects in an ordered medium, i.e., structures that cannot be fully removed by local and continuous deformations of the medium, have attracted persistent scientific fascination \cite{volovik:1977, Mermin:1979, Preskill:1991, Mineev:1998, Ueda:2014a}, including investigations of the compatibility of cosmic microwave background with the existence of topological defects in the early universe \cite{cruz:2007, feeney:2012} and the recent spatially resolved observations of monopoles \cite{IsolatedMonopole15}, knots \cite{Hall:2016}, and three-dimensional skyrmions \cite{Tiurev:2018} in quantum fields. Usually topological defects are treated as classical objects, but their quantum nature has attracted some attention recently \cite{dasgupta:2020}.

It has been understood since the late 1970s that topological defects can be classified using homotopy theory \cite{Mermin:1979, Michel:1980, Trebin:1982, Mineev:1998}, and that a fundamental role in the classification is played by the homotopy groups $\pi_n(X)$ of the order parameter space $X$, which is the space of all physically distinct local configurations of the system. For example, in a three-dimensional system, a nontrivial fundamental group $\pi_1(X)$ permits topologically distinct line defects, \emph{vortices}, and a nontrivial second homotopy group $\pi_2(X)$ permits topologically distinct point defects, \emph{monopoles}. However, in order to fully classify topological defect configurations and to study interactions between multiple defects, the homotopy groups of $X$ are not enough. An immediate manifestation of this is the behavior of monopoles around line defects: depending on the type of the monopole and the vortex, it may be possible that the charge of the monopole is altered as it travels around the singular line, or through a ring \cite{Ueda:2012, Ueda:2014b, Preskill:1991, Preskill:1992}. There exist condensed-matter realizations of these type of line defects, namely Alice strings and Alice rings that can appear in certain gauge field theories \cite{Preskill:1991, Preskill:1992, HEP:strings} and are thought to be a potential solution to the baryogenesis problem \citep{HEP:baryogenesis}. Remarkably, Alice rings have recently been observed in condensed matter systems \cite{AliceRing}, which may allow in the future experimental studies of the topology of these peculiar structures.

However, the existing literature is currently lacking clarity in the formulation of topological charge ambiguity and topological charge addition. When the order parameter space is real projective plane $\Rb \mathrm{P}^2$, which is also referred to as the space of nematic vectors,  these phenomena are well understood \cite{volovik:1977}. Namely, isolated charges are not classified by their integer charge $\pi_2(\RP^2) \cong \Zb$ but rather by its absolute value, due to the action of the fundamental group on $\pi_2$. However, in the presence of multiple charges, also the relative signs are well defined, leaving only the total sign of the system ambiguous. Moreover, the combined charge of several monopoles can be computed by locally orienting the directors around the coalescence path, leading to more refined information than charge combination using the orbit group of Trebin \cite{Trebin:1982}, which loses all information of the charges expect their values modulo 2.

In this article, we propose a conceptually clear description of charge ambiguity and the charge addition process using a generalization of the above picture. The mathematical framework of covering spaces and lifting results \cite{Hatcher} is well suited for this purpose, the case of nematic fields and their orientations corresponding to the two-sheeted covering of $\RP^2$ by the sphere $S^2$. As with nematics, this approach to charge addition allows us to retain more information than when employing the orbit group. The idea is to replace the order parameter space $X$ with a particularly symmetric covering space $\wtil X$ having the property that the topological charges of monopoles in fields taking values in $\wtil X$ are unambiguous, and can be added using the group law of $\pi_2(\wtil X)$.
 
In addition, we classify the topological defect configurations consisting of ring defects and monopoles, up to continuous deformations that leave the cores fixed. This is mathematically equivalent to classification up to continuous deformations that, roughly speaking, do not alter the topology of the core configuration \cite{HirschIsotopy}. The result is a generalization of the well-known classification of monopole configurations \cite{volovik:1977} and isolated ring defects \cite{Nakanishi:1988}. Classification in the presence of knotted cores is more involved \cite{machon:2014}, and we leave general investigations of knotted defects for future work. Instead, we advance the study of multiply charged monopoles by showing that such a monopole can be split into several monopoles of charge $\pm 1$ using an arbitrarily small local modification. Moreover, under certain physically realistic hypotheses, almost any small perturbation will have the same effect.

Throughout the work, we denote by $(X,x)$ the order parameter space considered as a \emph{pointed space}, i.e., $X$ is a topological space and $x \in X$ is a distinguished point, the \emph{basepoint}. In Section \ref{ChargeAmbAddSect} we study charge ambiguity and the process of charge addition using a particular covering space of $X$ as the main tool. In Section \ref{ClassificationSect} we classify topological defect configurations consisting of monopoles and ring defects by the means of elementary homotopy theory, and in Section \ref{SplittingSect} we study how multiply charged monopoles split using topology and differential geometry of manifolds. The Appendix consists of two sections. In Appendix \ref{CoveringSect}, we summarize the theory of covering spaces and deck transformations, and in Appendix \ref{MappingSpaceSect} we define mapping spaces and recall their basic properties. When referring to the Appendix, we refer to the exact position of the result being used. For example, the computation of the set of homotopy classes of maps in terms of the set of basepoint preserving homotopy classes of pointed maps will be referred to as Corollary \ref{PointedVsUnpointedClassCor}.


\section{Charge ambiguity and charge addition}\label{ChargeAmbAddSect}


In this section, we study charge ambiguity and charge addition using a particularly well-behaved covering space of the order parameter space $X$. This covering space allows us to enhance order parameter fields into closely related fields without charge ambiguity and where the combined charge of several monopoles can be computed using the group operation of the second homotopy group. Section \ref{SimplyConnectedSubSect} treats systems containing monopoles and Section \ref{MoreGeneralCaseSubSect} treats systems containing monopoles and ring defects.

Recall that a covering space $q: \wtil Y \to Y$  is a particularly simple finite-to-one map, and that the group $\Deck(\wtil Y / Y)$ of \emph{deck transformations} is the group of continuous symmetries of $\wtil Y$ compatible with $q$ in the sense that the symmetries preserve the preimages of $q$ (Definition \ref{DeckTransDef}). If both $Y$ and $\wtil Y$ are path connected spaces, then $q$ is a \emph{Galois covering} if it is maximally symmetric, i.e., if $\wtil Y / \Deck(\wtil Y / Y) = Y$ (Definition \ref{GaloisCoverDef}). For the rest of this section, we fix a pointed Galois covering $p: (\wtil X, \tilde x) \to (X, x)$ having the property that the $\pi_1(\wtil X, \tilde x)$-action on $\pi_2(\wtil X, \tilde x)$ (Definition \ref{FundGpdActionDef}) is trivial. By a special case of Proposition \ref{SmallestGaloisCoverProp}, there exists a smallest such covering, and this covering is unique up to a unique isomorphism of pointed covering spaces. A concrete example is given by the map from the two-sphere $S^2$ to the real projective plane $\RP^2$ that identifies antipodal points: $\Deck(S^2 / \RP^2) \cong \Zb_2$ is generated by the antipodal map and the map $S^2 \to \RP^2$ is clearly a Galois covering. Moreover, it is elementary that this is the smallest covering of $\RP^2$ trivializing the $\pi_1$-action on $\pi_2$. The reader may consult Table \ref{OPSpaceTable} to find other relevant examples of $X$ and $p$. 

We model the spatial region of our physical system, \emph{the cloud}, with the solid ball $\Bb^3 := \{ y \in \Rb^3 : \abs{\abs{y}} \leq 1 \}$, and denote by $C \subset \Bb^3$ the the points where the order parameter is well-defined. Concretely, $\Bb^3$ can be the extent of a spinor Bose--Einstein condensate, and $C$ the complement of the cores of the various singular defects contained in the cloud. The \emph{order parameter field} is modeled by a continuous map $\Psi: C \to X$. Given topological spaces $Y$ and $Z$, the set of homotopy classes of continuous maps $Y \to Z$ is denoted by $[Y, Z]$. 

\subsection{The simply connected case}\label{SimplyConnectedSubSect}

In this section, we restrict to the case where the order parameter field has finitely many monopoles and no other defects. In other words, $C$ is a solid ball $\Bb^3$ with finitely many points removed from its interior. Thus $C$ is simply connected. Given an embedded sphere $\Omega \subset C$, the finest homotopy theoretic invariant of $\Psi$ one can associate to $\Omega$ is the free homotopy class $[\Psi \vert_{\Omega}] \in [S^2, X]$, which by definition classifies the the winding of $\Psi$ along $\Omega$ up to homotopy. We call this element the \emph{type} of the \emph{topological charge contained inside} $\Omega$. As the free homotopy classes of maps from $S^2$ to $X$ correspond to $\pi_1(X,x)$-orbits in $\pi_2(X,x)$ \cite{Mermin:1979}, it makes sense to define the \emph{topological charge inside $\Omega$} as an element of $\pi_2(X,x)$ belonging to the $\pi_1(X,x)$-orbit corresponding to $[\Psi \vert_{\Omega}]$. The topological charge is by definition ambiguous up to the $\pi_1$-action on $\pi_2(X,x)$. If $\Omega$ encloses a single monopole at $y \in \Bb^3 \backslash C$, then the topological charge inside $\Omega$, and its type, are referred to as the \emph{topological charge of the monopole at $y$}, and the \emph{type of the monopole at $y$}, respectively. More generally, if $\Omega$ encloses the monopoles at $y_1, ..., y_n \in \Bb^3 \backslash C$ and no others, then the topological charge inside $\Omega$ is the \emph{combined topological charge of the monopoles at $y_1,...,y_n$}. The type of the combined charge is a well-defined element of $\pi_2(X,x) / \pi_1(X,x)$ rather than just of the orbit group \cite{Trebin:1982}, and therefore our approach retains more information than the one employing the orbit group.

 
 
  
  

\begin{figure}[h!]
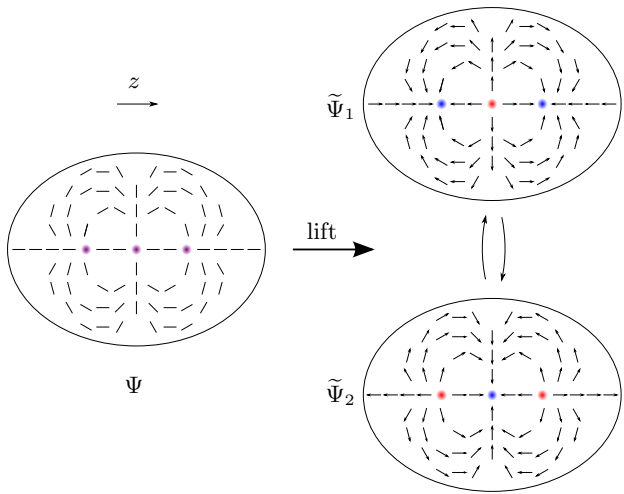

\include{Figures/1/figure}
\caption{Charge ambiguity and addition for a nematic vector field. A nematic field $\Psi$ taking values in $\Rb \mathrm{P}^2$ and its lifts $\wtil \Psi_1$ and $\wtil \Psi_2$ are shown. All configurations have $z$ as a rotational axis of symmetry. The field $\Psi$ is visualized as a field of arrows without heads, or rods, and the lifted fields $\wtil \Psi_1$ and $\wtil \Psi_2$, taking values in the universal cover $S^2$, are illustrated as the oriented versions of $\Psi$. Topological charges of the monopoles are well defined in the lifted fields $\wtil \Psi_i$, red and blue color indicating charge $+1$ and $-1$, respectively. The two lifts are interchanged by the action of $\Zb_2 \cong \Deck(S^2 / \RP^2)$ on $S^2$, which flips the orientations of the arrows and, as a result, the signs of the monopoles. Any two monopoles will have either the same sign or the opposite sign regardless of the chosen lift and the absolute value of the total charge inside the cloud is always $+1$.
}\label{ChargeAmbAddFig}
\end{figure}

In order to study these phenomena in more detail, we choose a continuous lift $\wtil \Psi: C \to \wtil X$, which exists by Lemma \ref{GeneralLiftingLem}. Because the $\pi_1$-action on $\pi_2(\wtil X, \tilde x) \cong \pi_2(X,x)$ is trivial, it follows that $[S^2, \wtil X]$ and $\pi_2(X,x)$ are isomorphic sets. Importantly, every monopole of the lifted field $\wtil \Psi$ has an unambiguous topological charge, and the charge addition process described in the previous paragraph reduces to the group law of $\pi_2(\wtil X, \tilde x) = \pi_2(X, x)$. Since $p: \wtil X \to X$ was assumed to be a Galois covering, the group $\Deck(\wtil X / X)$ of deck transformations (Definition \ref{DeckTransDef}) acts freely and transitively on the set of lifts $\wtil \Psi$ of $\Psi$. Moreover, Proposition \ref{GaloisGroupProp} provides an isomorphism $\Deck(\wtil X / X) \cong \pi_1(X,x)/\pi_1(\wtil X, \tilde x)$, which also identifies the action of $\Deck(\wtil X / X)$ on $[S^2, \wtil X]$ with the usual $\pi_1$-action on $\pi_2(X, x)$ (Remark \ref{DeckActionOnHtyRem}). Hence, the charge of a single monopole, computed in various lifts $\wtil \Psi$ of $\Psi$, ranges through all the charges in its type, as expected. A subtle point is that, even if two monopoles of the system are of the same type, they might lift to monopoles of different charges. If two monopoles lift to monopoles of different charge in one lift, then the same is true for all lifts. This behavior is illuminated by a concrete example in Figure \ref{ChargeAmbAddFig}.



\subsection{Beyond the simply connected case}\label{MoreGeneralCaseSubSect}

\begin{figure}[h!]
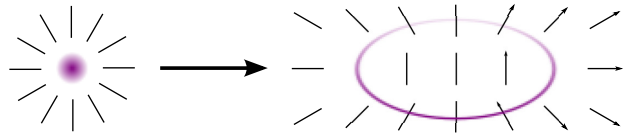

\include{Figures/2/figure}
\caption{Once a monopole in the nematic field decays into an Alice ring, it might become impossible to find lifts of the order parameter field. This failure is caused by the existence of closed loops that would have to flip the orientation, as illustrated in the figure. Mathematically, this is the content of Lemma  \ref{GeneralLiftingLem} and Remark \ref{InjectivityRem}.}\label{NoLiftFig}
\end{figure}

\begin{figure}[h!]
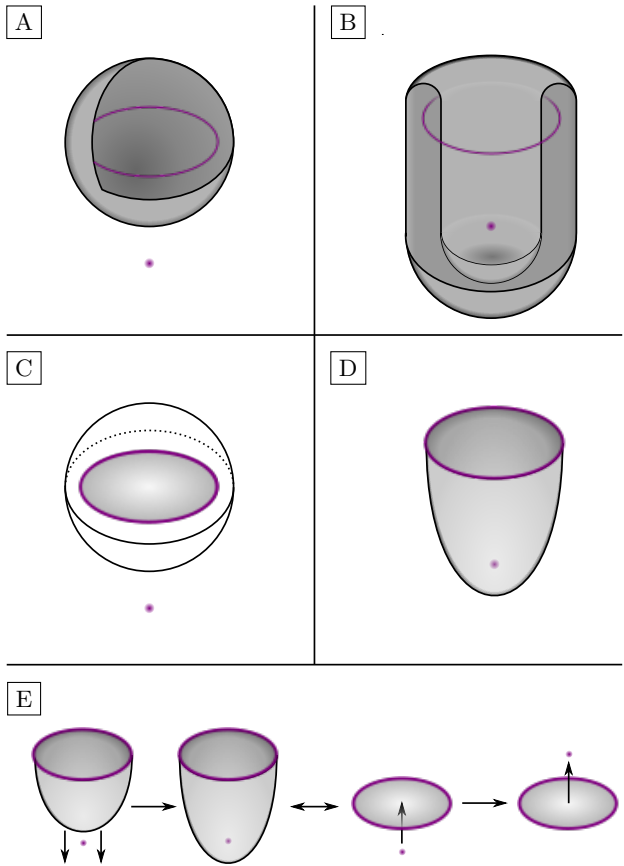

\include{Figures/3/figure}
\caption{Pictures (A) and (B) show two homotopically nonequivalent ways of enclosing the ring shaped singularity inside a sphere while leaving the monopole outside. If the ring is equipped with a membrane, as in (C), then there exists a unique homotopy class of embedded spheres that do not cross the membrane and enclose only the ring defect (C). Pictures (C) and (D) also depict homotopically non-equivalent choices of membranes. (E) The following processes are topologically equivalent: the membrane is moved past a monopole, the monopole is moved through the membrane. Note that the Cheshire charge of the ring after the monopole has passed through it is equal to the topological charge inside the sphere in (B). 
}\label{ChesireChargeFig}
\end{figure}

In this section, we consider an order parameter field that has, in addition to monopoles, singularities in the form of unknotted circles in the interior of the cloud. Such systems occur in condensed-matter physics, for example, if monopoles decay into Alice rings \cite{Ruostekoski:2003, AliceRing}. This situation creates a new complication: it might be impossible to lift the order parameter field, as is illustrated by Figure \ref{NoLiftFig}. However, one can define the $\pi_1$-equivalence class of topological charge of the field $\Psi$ inside an embedded sphere $\Omega \subset C$ as above, generalizing the definition of topological charge of an isolated monopole to this situation. We would like to define the topological charge of a ring defect in a similar fashion, but here a problem arises: there may be multiple homotopically inequivalent embedded spheres $S^2 \subset C$ that enclose only a single ring defect and no other defects---see Figure \ref{ChesireChargeFig} (A) and (B). Hence, without any further choices, the topological charge of a ring defect is ill defined, even as an element of $\pi_2(X,x)/\pi_1(X,x)$.

\begin{figure}[h!]
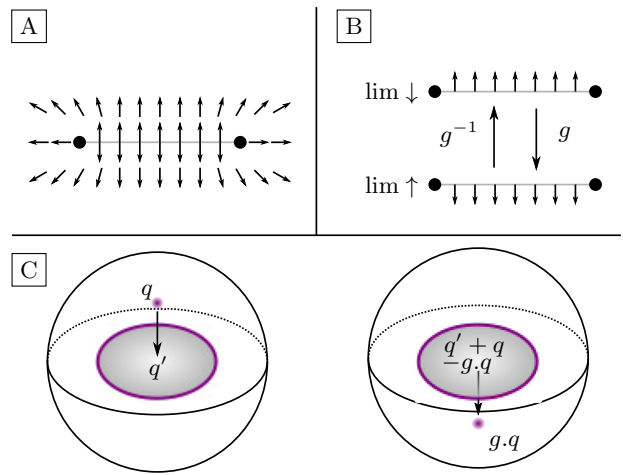

\include{Figures/4/figure}
\caption{(A) Slice of a ring-shaped singularity (purple dots) with a membrane (gray line) is shown together with the lifted order parameter field on the complement of the membrane. (B) The lifted order parameter field has two limiting configurations on the membrane that are interchanged by inverse deck transformations. (C) If a monopole passes through such a membrane, one has to apply $g$ to the order parameter field around it, altering also the topological charge of the monopole from $q$ to $g.q$. Since the total charge inside the sphere remains unchanged, the Cheshire charge of the ring must change by $q - g.q$ \cite{Preskill:1991, Ueda:2014b}. 
}\label{ChargeTransferFig}
\end{figure}

A conceptually clear solution to all of the problems mentioned above is provided by equipping each singular ring with a \emph{membrane}, that is, a two-dimensional disk, possibly deformed, closing the loop. The complement $C'$ of the membranes in $C$ is simply connected. Moreover, the ring-shaped core has a unique homotopy class of enclosing spheres that do not cross the membrane, and the topological charge inside such a sphere is referred to as the \emph{Cheshire charge} of the ring defect \cite{Preskill:1991, Preskill:1992}. Importantly, the restriction $\Psi'$ of $\Psi$ to $C'$ admits a lift $\wtil \Psi': C' \to \wtil X$ by Lemma \ref{GeneralLiftingLem}, and the lifted field $\wtil\Psi'$ behaves in an interesting fashion near the membranes. Clearly, there are two limiting configurations on the membrane obtained by approaching it from opposite sides (Figure \ref{ChargeTransferFig} (B)). These membranes are analogous to the gauge discontinuities that appear in the study of Alice rings in cosmology \cite{Preskill:1991, Preskill:1992}, and branch cuts that appear when investigating knotted defects in nematic liquid crystals \cite{machon:2014}. As each membrane is connected, because $\Psi$ is defined on the membrane, and because $p: \wtil X \to X$ is a Galois covering, there exists a unique pair of inverse deck transformations interchanging the limiting configurations (Figure \ref{ChargeTransferFig} (C)). A convenient, although unphysical, way to mathematically describe this behavior is to consider every ring with a membrane as a portal that operates on fields passing through by applying to them a discrete symmetry of $\wtil X$. As above, $\Deck(\wtil X / X)$ acts freely and transitively on the set of lifts $\wtil \Psi'$. The effect of this action on the membranes is fairly simple: it transforms the associated deck transformations according to the conjugation action of $\Deck(\wtil X / X)$ on itself.


The framework described in the previous paragraph also provides some insight on \emph{charge transfer} which has been studied previously \cite{Ueda:2012, Ueda:2014a, Ueda:2014b}. The order parameter field near a monopole which travels through a membrane is transformed by the deck transformation $g$ associated to it. This can affect the charge of the traveling defect, and consequently charge conservation dictates that that the Cheshire charge of the ring must also be affected, as shown in Figure \ref{ChargeTransferFig}.

We end this section by proposing a mathematical definition of an Alice ring which captures the essential topological features. The proposed definition generalizes cosmic Alice rings \cite{Preskill:1991, Preskill:1992} and half-quantum vortex rings \cite{Ueda:2012, Ueda:2014b}.

\begin{defn}
Let $D \subset \Bb^3$ be an unknotted ring-shaped singularity, and let $\gamma$ be a small loop that winds about the singularity once. We define $D$ to be an \emph{Alice ring (for monopoles)} if any element, or equivalently every element, of the conjugacy class of $\pi_1(X,x)$ corresponding to $\Psi \circ \gamma : S^1 \to X$ acts nontrivially on $\pi_2(X,x)$. In physical terms, $D$ is an Alice ring if it there exists a type of monopoles, the charge of which would be altered by traveling about $D$.
\end{defn}

\begin{table*}[t]
	\caption{Examples of physically relevant order parameter spaces, their first and second homotopy groups and the $\pi_1$-action on $\pi_2$. The target of the covering is the order parameter space of the system, and the source is the smallest Galois covering space trivializing the $\pi_1$-action on $\pi_2$. The $\pi_1$-action on $\pi_2$ is denoted by $\alpha . \beta$, where $\alpha \in \pi_1$ and $\beta \in \pi_2$. The last example is that of a grand unified theory (GUT) equipped with a Higgs field breaking the gauge group $\SO(10)$ down to $\SO(2)^5 \rtimes S_5$, where the symmetric group $S_5$ acts on $\SO(2)^5$ by permuting the factors, and where $\Zb^5_\mathrm{even} \subset \Zb^5$ consists of such $5$-tuples $(m_1,...,m_5)$ that $m_1 + \cdots + m_5$ is even. Since $\SO(2)^5 \rtimes S_5$ is a closed subgroup of $\SO(10)$, there exists a continuous $\SO(10)$-representation $V$ (\emph{Higgs field}) and a vector $v \in V$ (\emph{vacuum vector}) producing the desired symmetry breaking \cite{Mostow1957, Palais1957}. This is a toy model: its purpose is to illustrate interesting topological behavior by means of a simple example, rather than provide a physically realistic example of a symmetry breaking pattern potentially useful in describing our universe. However, it should be possible to construct more realistic models exhibiting interesting $\pi_1$-actions on $\pi_2$ using the same idea.
	}\label{OPSpaceTable}
	
	\begin{center}
	\begin{tabular}{ | l | l | l | l | l | l | }
    \hline
    system  & phase & covering space & $\pi_1$ & $\pi_2$  & $\pi_1$-action \\ \hline
    liquid crystal & UN \cite{Ueda:2012} & $S^2 \to \RP^2$ & $\Zb_2$ & $\Zb$ & $[n].m = (-1)^n m$  \\
    gaseous BEC & spin-1 polar \cite{Ueda:2012} & $S^1 \times S^2 \to (S^1 \times S^2)/\Zb_2$ & $\Zb$ & $\Zb$ & $n.m = (-1)^n m$  \\ 
	& spin-2 UN \cite{Ueda:2012} & $S^1 \times S^2 \to S^1 \times \RP^2$ & $\Zb \times \Zb_2$ & $\Zb$ & $(n_1, [n_2]).m = (-1)^{n_2} m$ \\
	$^3$He-A & dipole free \cite{Ueda:2012} & $S^2 \times \SO(3) \to (S^2 \times \SO(3))/\Zb_2$ & $\Zb_4$ & $\Zb$ & $[n].m = (-1)^n m$  \\
	$\SO(10)$ GUT & various & $\SO(10) / \SO(2)^{5} \to \SO(10) / \big[ \SO(2)^{5} \rtimes S_5 \big]$ & $S_5$ & $\Zb^{5}_\mathrm{even}$ & $(\sigma, [n]) . (m_1,...,m_5) = (m_{\sigma_1}, ..., m_{\sigma_5})$ \\
    \hline
    \end{tabular}
	\end{center}
\end{table*}

\section{Configurations of monopoles and ring defects}\label{ClassificationSect}

In this section, we study and classify defect configurations consisting of monopoles and unknotted ring defects, generalizing the well-known classification of configuration of monopoles \cite{volovik:1977} and of isolated ring defects \cite{Nakanishi:1988}. Our method is to decompose the complement of the cores into simple pieces, and then invoke standard properties of mapping spaces reviewed in Appendix \ref{MappingSpaceSect}. 

More precisely, we aim to find the elements of the set $[C, X]$, where $C$---as in Section \ref{MoreGeneralCaseSubSect}---is the complement of isolated points and unknotted circles inside a solid ball $\Bb^3$. Note that the elements of $[C, X]$ are the \emph{topological defect configurations} of the system with a prespecified core configuration, considered up to continuous deformations leaving the cores fixed. However, as any smooth isotopy of the configuration of cores---roughly speaking, a smooth deformation in which defect cores are not allowed to be pinched, to cross each other, or to be pierced by the order parameter field---can be canceled by an \emph{ambient isotopy} of $\Bb^3$ \cite{HirschIsotopy}, the set $[C, X]$ may be regarded as classifying topological defect configurations up to continuous deformations that change the core configuration up to smooth isotopy. 

We recall the \emph{wedge summation} of pointed topological spaces. This is the operation that, provided with two pointed spaces $(Y,y)$ and $(Z,z)$, produces the pointed space $(Y \lor Z, *)$, where $Y \lor Z$ is the space obtained by attaching $Y$ and $Z$ along $y$ and $z$, and $* \in Y \lor Z$ is the point of attachment. The defining feature of this construction is that, for any third pointed space $(W,w)$, there exists a canonical homeomorphism
\begin{equation}\label{WedgeSumEq}
\Map_*(Y \lor Z, W) \cong \Map_*(Y, W) \times \Map_*(Z, W),
\end{equation}
where $\Map_*$ denotes the space of basepoint-preserving continuous maps (Definition \ref{MappingSpaceDef}). As a consequence,
\begin{equation}\label{WedgeSumHomEq}
[Y \lor Z, W]_* \cong [Y, W]_* \times [Z, W]_*,
\end{equation}
where $[A,B]_*$ stands for the basepoint-preserving homotopy classes of maps of pointed spaces from $A$ to $B$.

\begin{figure}[h!]
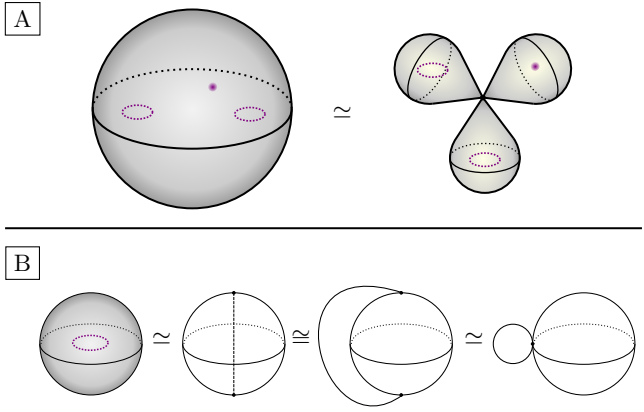

\include{Figures/5/figure}
\caption{
(A) Solid ball containing two unknotted ring defects and a monopole (purple dashed circle, purple dot) is equivalent to the wedge sum of several balls containing a single defect. (B) The complement of an unknotted circle inside a solid ball is equivalent to the wedge sum $S^1 \vee S^2$. The intermediate spaces are a hollow sphere with a chord running through it, and the same (abstract) topological space presented by a different embedding.
}\label{ComplementFig}
\end{figure}

As illustrated in Figure \ref{ComplementFig}, $C$ is homotopically equivalent to a wedge sum of several clouds, each of which contains a single defect. Each of these pieces is equivalent to either $S^2$ if the cloud contains a monopole or $S^1 \lor S^2$ if the cloud contains a ring defect. Thus 
\begin{equation}\label{CloudDecompositionEq}
C \cong \Bigg( \bigvee_y S^2_y \Bigg) \vee \Bigg( \bigvee_d S^1_d \vee S^2_d \Bigg),
\end{equation}
where $y$ and $d$ range over all the monopoles and the ring defects of the system, respectively. Applying (\ref{WedgeSumHomEq}) and the definition of homotopy groups, we identify the set of pointed homotopy classes as
\begin{equation}\label{PointedClassesEq}
[C, X]_* \cong \bigg[ \prod_y \pi_2(X,x)_y \bigg] \times \bigg[ \prod_d  \pi_1(X,x)_d \times \pi_2(X,x)_d \bigg].
\end{equation}
By Corollary \ref{PointedVsUnpointedClassCor}, the free homotopy classes can be computed as
\begin{equation}\label{FreeClassesEq}
[C, X] \cong [C,X]_* / \pi_1(X,x),
\end{equation}
where $\pi_1(X,x)$ acts on each factor on the right side of (\ref{PointedClassesEq}) by the usual $\pi_1$-action on $\pi_i$ (Definition \ref{FundGpdActionDef}). In other words, the equivalence class of the configuration is completely determined by the topological charges of all the defects and the topological vorticities of all the ring defects, and these data are ambiguous up to the simultaneous action of $\pi_1(X,x)$.

\section{Instability of multiply charged monopoles}\label{SplittingSect}

In this section, we study the stability properties of isolated monopoles. We will proceed by identifying the topological charge with the degree of a proper map of manifolds and then invoking well-known transversality theorems. The physical interpretation of this is that a multiply charged monopole may be split into multiple singly charged monopoles.

As in Section \ref{SimplyConnectedSubSect}, $C$ is the complement of finitely many points in the interior of a solid ball. We further assume that the order parameter space is one of the examples in Table \ref{OPSpaceTable} the covering space $\wtil X$ of which is of the form $S^2 \times Y$, where $Y$ is a space with trivial $\pi_2(Y)$. An order parameter field $\Psi: C \to X$ admits a lift $\wtil \Psi: C \to S^2 \times Y$ by Lemma \ref{GeneralLiftingLem}, by neglecting the $Y$-component of $\wtil \Psi$, and accounting for the magnitude of the order parameter, we obtain a map $\Phi: \Bb^3 \to \Rb^3$. Note that $C$ is the preimage of $\Rb^3 \backslash \{\bar 0\}$, where $\bar 0$ denotes the origin, and monopoles are exactly the points mapping to $\bar 0$. A monopole at $y \in \Bb^3$ is \emph{regular} if $\Phi$ is a local homeomorphism at $y$, i.e., there exists an open neighborhood of $y$ mapping homeomorphically onto an open neighborhood of $\bar 0$. Such a monopole has degree $\pm 1$. We assume that none of the points of the boundary $\partial \Bb^3$ are sent to $\bar 0$. 

Let $U$ be an open ball around $\bar 0 \in \Rb^3$ that is small enough to be contained in the complement of $\Phi (\partial \Bb^3)$, and assume for simplicity that $V := \Phi^{-1} U$ is homeomorphic to the disjoint union $\coprod_{i \in I} V_i$, where each $V_i$ contains a single monopole of degree $n_i$. Since the restrictions $\Phi_i: V_i \to U$ are \emph{proper maps}---in other words, preimages of compact sets are compact---they have well-defined \emph{Brouwer degrees} \cite{Milnor:1965, Epstein:1966, Hirsch:1976}, which in this case are equal to $n_i$. Hence, the monopoles inside $V_i$ can be split into exactly $\abs{n_i}$ regular monopoles by arbitrarily small local modifications to $\Phi_i$ \cite{Epstein:1966}. We conclude that a multiply charged monopole can always be split into several singly charged monopoles using only a small local perturbation.

\begin{figure}[h!]
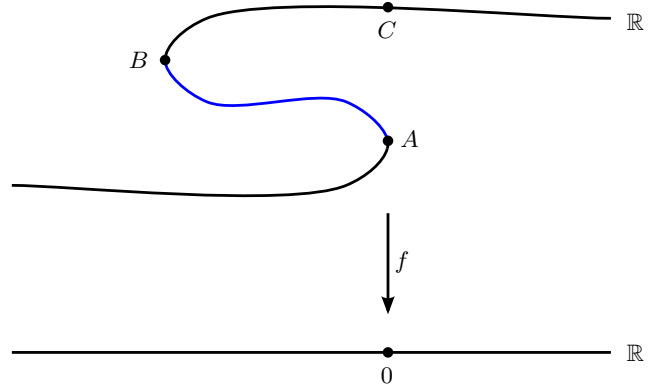

\include{Figures/6/figure}
\caption{ 
A proper and continuously differentiable map $f: \Rb \to \Rb$. The map is orientation reversing on the segment between points $A$ and $B$ (marked in blue), and orientation preserving elsewhere. It is not open at $A$ and $B$ where it changes orientation, as the images of small open neighborhoods of these points are half-open intervals. If we regard $f$ as an order parameter field, there exists a ``monopole'' of degree 1 at $C$ and a monopole of degree 0 at $A$. A small perturbation either completely destroys the monopole at $A$ or creates a $\pm 1$ monopole-antimonopole pair in its place, depending on the direction of the perturbation. In either case, the total charge is conserved.
}\label{FoldFig}
\end{figure}

In fact, multiply charged monopoles are extremely unstable in the following sense. Let us call  $v \in V_i$ a \emph{regular point} of $\Phi_i$ if $\Phi_i$ is a local homeomorphism at $v$, and $u \in U$ a \emph{regular value} of $\Phi_i$ if $\Phi_i^{-1} \{u\}$ consists of finitely many regular points. A point or a value that is not regular is called \emph{critical}. If we assume that $\Phi_i : V_i \to U$ is \emph{open} and \emph{discrete} \cite{Vaisala:1966}, sets of critical points and critical values have a codimension of 2 \cite{Vaisala:1966, Chernavskii:1964}. It follows that for every regular value $u$, i.e., every point of $U$ outside a codimension $2$ subset, the preimage of $u$ consists of $\abs{n_i}$ regular points. If we assume $\Phi_i$ to be once continuously differentiable instead, then by Sard's theorem \cite{Milnor:1965, Sard:1942} the set of critical values of $\Phi_i$ has measure $0$, and so the preimage of almost any $u \in U$ consists of regular points, whose degrees add up to $n_i$---see Figure \ref{FoldFig} for illustration of this behavior in one dimension. Hence, in either of the two cases considered above, applying almost any small constant perturbation to $\Phi$ will split all the monopoles into monopoles of degrees $\pm 1$ in a way that preserves the total degree.

\section{Conclusions}

We studied charges of topological defects using a particularly symmetric covering space of the order parameter space $X$. Even though the charge of an isolated monopole is well defined only up to the $\pi_1$-action on $\pi_2(X)$, the configuration of $n$ charges is well-defined up to a simultaneous $\pi_1$-action on $\prod_{i=1}^n\pi_2(X)$, allowing for the computation of the combined charge of several monopoles as an element of $\pi_2(X)/\pi_1(X)$. This is an improvement over the orbit group approach \cite{Trebin:1982} as more information is retained. Our method also illuminates the influence of ring-shaped defects on monopoles and the phenomenon of Cheshire charge. We also suggested a definition for an Alice ring in purely topological terms: it is a ring defect whose vorticity, considered as an element of $\pi_1$, has a nontrivial action on $\pi_2$.

We classified defect configurations consisting of monopoles and ring defects, up to continuous deformations that leave the cores fixed. The configurations are classified by the topological charges of the defects and  the vorticities around the cores of the ring defects, up to simultaneous $\pi_1$-actions. We also studied the stability of multiply charged monopoles and showed that it is possible to split a multiply charged monopole into monopoles of degrees $\pm 1$ using only small local modifications to the order parameter field. In physical terms this implies that a multiply charged monopole is topologically equivalent to a defect configuration consisting of multiple singly charged monopoles.

\begin{acknowledgments}
We have received funding from the European Research Council under Grant No 681311 (QUESS) and from the Academy of Finland Centre of Excellence program (project 336810). The authors would like to thank Alina Blinova, David Hall, Valtteri Lahtinen and Roberto Zamora-Zamora for useful comments.
\end{acknowledgments}

\appendix

\section{Covering spaces}\label{CoveringSect}

In this section, we recall the basics of the theory of covering spaces. A basic reference with many pictures is Chapter 1.3 of Hatcher's book \cite{Hatcher}. As a general rule, we give full statements of the results, but we do not explain the technical terminology unless absolutely necessary. We stress here that all the technical assumptions appearing in this section hold for manifolds, which should be connected if the result asks for connectivity. We do not repeat repeat the statement of this assumption below.

\begin{defn}
Let $X$ be a topological space. Then a \emph{covering space} of $X$ is a continuous morphism $p: \wtil X \to X$ having the property that for any $x \in X$, there exists an open neighborhood $U$ of $x$ and a homeomorphism $\phi: \coprod_{i \in I} U_i \to p^{-1}U$---each $U_i$ being a copy of $U$---such that the composition
$$\coprod_{i \in I} U_i \xrightarrow{\phi} p^{-1}U \xrightarrow{p \vert_{p^{-1}U}}  U$$
restricts to the identity map on each $U_i$. If $X$ is path connected, then the \emph{degree} $\deg(p)$ of $p$ is the cardinality of any set $I$ that appears as above. 
\end{defn}

\begin{lem}[Path lifting lemma]\label{PathLiftingLem}
Let $p: \wtil X \to X$ be a covering space, and let $\gamma: [0,1] \to X$ be a continuous path in $X$. Let $\tilde x_0 \in \wtil X$ be such that $p(\tilde x_0) = \gamma(0)$. Then there exists a unique continuous map $\tilde \gamma$ such that $\gamma = p \circ \tilde \gamma$ and $\tilde \gamma(0) = \tilde x_0$. The path $\tilde \gamma$ is called the \emph{lift of $\gamma$ starting at $\tilde x_0$}. 

Analogously, for each $\tilde x_1$ satisfying $p(\tilde x_1) = \gamma(1)$, there exists a unique \emph{lift of $\gamma$ ending at $\tilde x_1$}.
\end{lem}

\begin{defn}\label{MonodromyDef}
Let $p: \wtil X \to X$ be a covering space, $x \in X$, and consider the \emph{fiber} over $x$, $F_x := p^{-1} \{x\}$. The \emph{monodromy action} of $\pi_1(X, x)$ on $F_x$ is defined as follows: given $[\gamma] \in \pi_1(X, x)$ and $a \in F_x$, we define $[\gamma].a$ as $\tilde\gamma(0),$ where $\tilde\gamma$ is the unique lift of $\gamma$ ending at $a$. The result does not depend on the chosen representative $\gamma$. 
\end{defn}

%

The path lifting lemma can be used to prove a more general lifting result. Before this, we need the following definition.

\begin{defn}\label{PointedSpaceDef}
A \emph{pointed space} is a pair $(X, x)$, where $X$ is a topological space and $x \in X$. A \emph{morphism} or a \emph{map} $f: (X, x) \to (Y, y)$ of pointed spaces is a continuous map $f: X \to Y$ satisfying $f(x)=y$. 
\end{defn}

\begin{lem}[General lifting lemma]\label{GeneralLiftingLem}
Let $p: (\wtil X, \tilde x) \to (X, x)$ be a (pointed) covering map, and let $f: (Z, z) \to (X, x)$ be a morphism of pointed spaces, where $Z$ is connected and locally path connected. Then there exists a unique \emph{lift} $\tilde f: (Z,z) \to (\wtil X, \tilde x)$---i.e., $f = p \circ \tilde f$---if and only if the image of $f_*: \pi_1(Z,z) \to \pi_1(X,x)$ is contained in the image of the injective homomorphism $p_*: \pi_1(\wtil X,\tilde x) \hook \pi_1(X,x)$.
\end{lem}

\begin{rem}\label{InjectivityRem}
Consider the fiber sequence
$$(F_x, \tilde x) \stackrel{i}{\hookrightarrow} (\wtil X, \tilde x) \xrightarrow{p} (X,x),$$
where $i$ is the inclusion of the fiber over $x$. The long exact homotopy sequence gives the following sequence of maps
$$\{e\} \to \pi_1(\wtil X, \tilde x) \xrightarrow{p_*} \pi_1(X,x) \xrightarrow{\delta} F_x,$$
proving immediately that $p_*$ is injective. The map $\delta$ is closely related to the monodromy action as $\delta$ sends $[\gamma]$ to $[\gamma].\tilde x$. The class $[\gamma]$ lies in the image of $p_*$ if and only if it lifts to a closed loop starting at $\tilde x$. If $\wtil X$ is path connected, $\delta$ is a surjection, and hence $F_x$---equipped the monodromy action---is isomorphic to the $\pi_1(X,x)$-set $\pi_1(\wtil X, \tilde x) \backslash \pi_1(X,x)$ (left cosets).
\end{rem}

\begin{defn}\label{DeckTransDef}
The group of the covering-space automorphisms of $p: \wtil X \to X$, i.e., homeomorphisms $\psi: \wtil X \to \wtil X$ such that $p = p \circ \psi$, is called the group of \emph{deck transformations} of $p$, and it is denoted by $\Deck(\wtil X/X)$. The canonical group action of $\Deck(\wtil X/X)$ on $\wtil X$ is compatible with the projection $p$ by definition. Hence, we obtain an induced morphism $\bar p: \wtil X / \Deck(\wtil X/X) \to X$. 
\end{defn}

\begin{defn}\label{GaloisCoverDef}
If both $\wtil X$ and $X$ are path connected and $\bar p$ is a homeomorphism, then $p$ is called a \emph{Galois covering}. Note that by Lemma \ref{GeneralLiftingLem}, the action of $\Deck(\wtil X/X)$ restricts to a free action on the fibers $F_x$. It is clear that $p$ is Galois if and only if the action is transitive as well. Moreover, if $\deg(p)$ is finite, then $p$ being a Galois covering is equivalent to the equality $\abs{\Deck(\wtil X/X)} = \deg(p)$.
\end{defn}

\begin{figure}[h!]
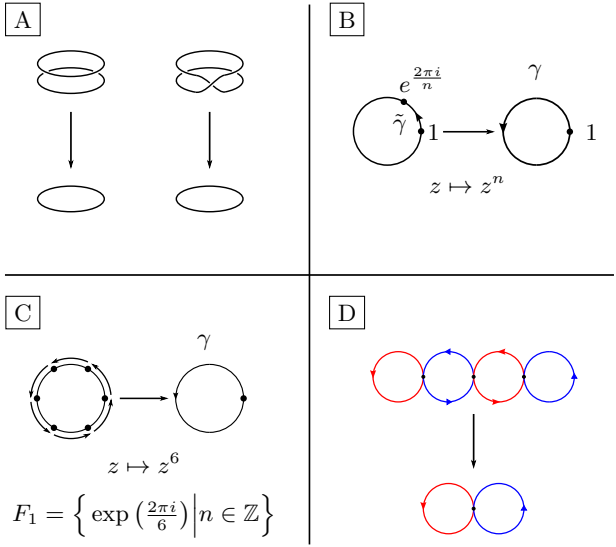

\include{Figures/7/figure}
\caption{(A) Examples of a trivial and nontrivial covering space (of degree 2) of the circle $S^1$. 
(B) Nontrivial covering of degree $n$ of $S^1$, regarded as the unit circle in the complex plane. The unique lift $\tilde \gamma$ of the path $\gamma$ starting at 1 is indicated in the picture. The group of deck transformations is given by rotations by $2 \pi/n$ radians. Since there are exactly $n$ such rotations this is a Galois covering. 
(C) The picture shows the fiber $F_1$ over $1$ of the 6-fold cover. The monodromy action of $[\gamma]$ cyclically permutes the elements in the fiber. Note that for each deck transformation $\psi$, there exists a homotopy class of loops whose monodromy action on $F_1$ coincides with the action of $\psi$ on $F_1$ (cf. Proposition \ref{GaloisCoverCharProp}). 
 (D) An example of a non-Galois covering space: the only deck transformation is the identity, but the covering has degree 3.}
\end{figure}

\begin{ex}[Finite Galois coverings arise as quotients]
If $X$ is a connected Hausdorff topological space and a finite group $G$ acts on it freely, then the quotient map $X \to X/G$ is a Galois covering with $\Deck(X / (X/G)) = G$. Conversely, if $p: \wtil X \to X$ is a Galois covering, then $p$ is equivalent to $\wtil X \to \wtil X / \Deck(\wtil X / X)$.
\end{ex}

\begin{prop}[Characterization of Galois covers]\label{GaloisCoverCharProp}
Let $X$ be connected and locally path connected, and suppose $p: \wtil X \to X$ is a connected covering. Then $p$ is a Galois covering if and only there exists such a $\tilde x \in \wtil X$ that the subgroup
$$p_*: \pi_1(\wtil X, \tilde x) \hook \pi_1(X, p(\tilde x))$$
is normal. It this is the case, then the above holds for all $\tilde x \in \wtil X$.
\end{prop}

\begin{prop}[Presentation of the Galois group]\label{GaloisGroupProp}
Let $(\wtil X, \tilde x) \to (X,x)$ be a pointed Galois covering. Since $\Deck(\wtil X / X)$ acts freely and transitively on $F_x$, the function 
$$\phi: g \in \Deck(\wtil X / X) \mapsto g.\tilde x \in F_x$$ 
is a bijection of sets. Thus the composition
$$\phi^{-1} \circ \delta: \pi_1(X,x) \to \Deck(\wtil X / X),$$
where $\delta$ is as in Remark \ref{InjectivityRem}, is a surjective group homomorphism with kernel $\pi_1(\wtil X, \tilde x)$. In particular 
$$\Deck(\wtil X / X) \cong \pi_1(X,x)/\pi_1(\wtil X, \tilde x).$$
\end{prop}

\begin{figure}[h!]
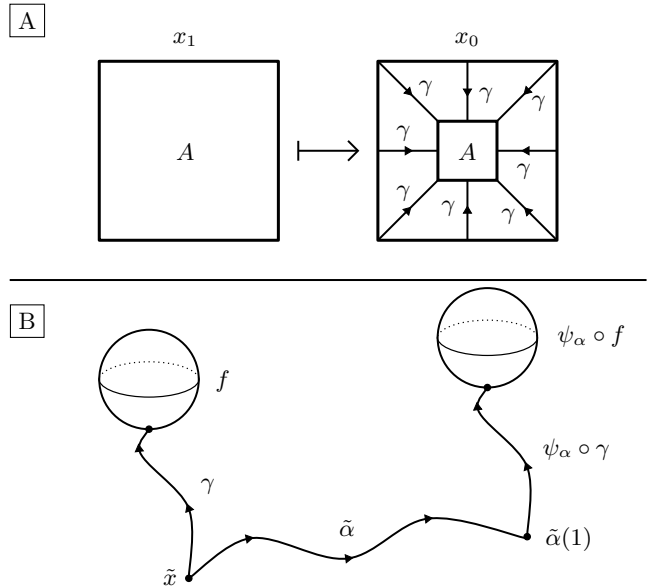

\include{Figures/8/figure}
\caption{(A) The action of a path $\gamma$ from $x_0$ to $x_1$ on higher homotopy groups. Left: an element $[A] \in \pi_n(X,x_1)$ is presented by depicting $A$ as a continuous map from the $n$-hypercube to $X$ taking constant value $x_1$ at the boundary. Right: the class $\gamma_*[A]$ is presented in a similar fashion. (B) The path $\wtil \alpha$ in $\wtil X$ is the unique lift of a closed loop $\alpha$ in $X$ starting at $\tilde x$. If there exists a deck transformation $\psi_\alpha$ satisfying $\psi_\alpha(\tilde x) = \tilde\alpha(1)$ (unique such deck transformation exists if $\wtil X \to X$ is a Galois covering, see Proposition \ref{GaloisGroupProp}), then is transforms a sphere $f: S^2 \to \wtil X$ attached to $\tilde x$ by a path $\gamma$ to a sphere that is attached in a similar manner to $\tilde{\alpha}(1)$.
}\label{HtyActionFig}
\end{figure}

\begin{defn}\label{FundGpdActionDef}
Let $X$ be a topological space. Then, a path $\gamma$ in $X$ with $\gamma(0) = x_0$ and $\gamma(1) = x_1$ induces group homomorphisms $\gamma_*: \pi_i(X, x_1) \to \pi_i(X, x_0)$ for all $i \geq 1$, see Figure \ref{HtyActionFig} for details. The homomorphism $\gamma_*$ depends only on the path homotopy class of $\gamma$, i.e., it does not change if $\gamma$ is deformed while keeping the endpoints fixed. In particular, for every $x \in X$, we obtain a group action of $\pi_1(X,x)$ on $\pi_i(X,x)$, where each $[\gamma] \in \pi_1(X,x)$ operates as $\gamma_*$ on $\pi_i(X,x)$. This action is referred to as the \emph{$\pi_1$-action} on the homotopy groups.  
\end{defn}

\begin{rem}[Action of $\Deck(\wtil X / X)$ is the $\pi_1$-action]\label{DeckActionOnHtyRem}
Let $p: (\wtil X, \tilde x) \to (X, x)$ be a pointed covering space, $i \geq 2$, and let $(S^i, e)$ be the $i$-sphere pointed at the North pole $e$. If the $\pi_1(\wtil X, \tilde x)$-action on $\pi_i(\wtil X, \tilde x)$ is trivial, the set $[S^i, \wtil X]$ of free homotopy classes of spheres is isomorphic to $\pi_i(\wtil X, \tilde x)$. Indeed, given a continuous map $f: S^2 \to \wtil X$, choose a path $\gamma$ from $\tilde x$ to $f(e)$ and let $A_f := \gamma_*([f]) \in \pi_i(X, \tilde x)$. The class $A_f$ does not depend on the choice of $\gamma$: given another path $\gamma'$ from $\tilde x$ to $f(e)$, then $\gamma_*([f])$ and $\gamma'_*([f])$ differ by the action of a closed loop on $\pi_i(\wtil X, \tilde x)$. Since the $\pi_1$-action was assumed to be trivial, $A_f$ does not depend on the choice of $\gamma$. We have described a bijection $\nu: [S^i, \wtil X] \to \pi_i(X,x)$.

Suppose then that $p$ is a Galois covering, and consider the deck transformation $\psi_\alpha$ associated to $[\alpha] \in \pi_1(X,x)$ as in Proposition \ref{GaloisGroupProp}. Then, as in Figure \ref{HtyActionFig} (B), $\psi_a \circ f: S^i \to \wtil X$ can be connected to $\tilde x$ by composing $\tilde\alpha$ with $\psi_\alpha \circ \gamma$, where $\tilde \alpha$ is the lift of $\alpha$ starting at $\tilde x$. If we denote by $B_f \in \pi_i(\wtil X, \tilde x)$ the element thus obtained, then
$$p_*(B_f) = \alpha_* [ p_*(A_f)] \in \pi_i(X,x),$$
where $p_*$ is the homomorphism $\pi_i(\wtil X, \tilde x) \to \pi_i(X,x)$. Hence,
$$(p_* \circ \nu) \big(\psi_\alpha [f]\big) = \alpha_*\big[(p_* \circ \nu)[f]\big],$$
identifying the $\Deck(\wtil X/X)$-action on $[S^i, \wtil X]$ with the $\pi_1$-action on $\pi_i(X,x)$.
\end{rem}

\begin{prop}\label{SmallestGaloisCoverProp}
Let $(X,x)$ be connected, locally path connected and semi-locally simply connected, and let $I \subset \Nb_{\geq 1}$. The $\pi_1$-action on the homotopy groups provides a group homomorphism
$$\pi_1(X,x) \to \prod_{i \in I} \mathrm{Aut}\big[\pi_i(X,x)\big],$$
where $\mathrm{Aut}(G)$ denotes the automorphism group of $G$. Let us denote the kernel of the above homomorphism by $K$. There exists a connected pointed covering space $(\wtil X, \tilde x)$ with $\pi_1(\wtil X, \tilde x) = K$, and this covering space is unique up to unique isomorphism of pointed covering spaces. Moreover, this is a Galois covering space as kernels are normal.

The map $p: \wtil X \to X$ is the smallest covering of $X$, on which the $\pi_1$-action on $\pi_i(\wtil X, \tilde x)$ is trivial for all $i \in I$: given another covering space $q: \wtil X' \to X$ with this property, there exists a continuous map $\psi: \wtil X' \to \wtil X$ satisfying $q = p \circ \psi$.
\end{prop}

\section{Mapping spaces}\label{MappingSpaceSect}

In this section, we recall the basics of the theory of mapping spaces. A basic reference is Chapter 5 of tom Dieck's book on algebraic topology \cite{tomDieck}. 

\begin{defn}\label{MappingSpaceDef}
Let $X$ and $Y$ be compactly generated spaces (e.g. unions of manifolds). The \emph{mapping space} $\Map(X,Y)$ is the set of all continuous maps $X \to Y$ equipped with the compact-open topology. If $(X,x)$ and $(Y,y)$ are compactly generated pointed spaces, then the \emph{pointed mapping space} $\Map_*(X,Y)$ is the subspace of $\Map(X,Y)$ consisting of those continuous morphisms $f: X \to Y$ satisfying $f(x) = y$.
\end{defn} 

\begin{prop}
Let $I$ denote the unit interval, and let $(X,x)$ and $(Y,y)$ be compactly generated pointed spaces. Then there exists a natural homeomorphism
$$\Map\big[I, \Map(X, Y)\big] \cong \Map(I \times X, Y)$$
that identifies the paths in $\Map(X, Y)$ with homotopies of continuous maps $X \to Y$. Similarly, the composition
\begin{align*}
\Map\big[I, \Map_*(X, Y)\big] &\to \Map\big[I, \Map(X, Y)\big] \\
&\cong \Map(I \times X, Y)
\end{align*}
identifies the paths in $\Map_*(X, Y)$ with basepoint-preserving homotopies of morphisms $(X,x) \to (Y,y)$. 
\end{prop}

The previous result identifies the set of path components of $\Map(X,Y)$ with the set $[X,Y]$ of homotopy classes of continuous maps $X \to Y$. Similarly, the path components of $\Map_*(X,Y)$ are identified with $[X,Y]_*$, that is the set of morphisms $(X,x) \to (Y,y)$ up to basepoint-preserving homotopies.

\begin{thm}
If $(X,x)$ and $(Y,y)$ are pointed spaces, and $(X,x)$ is well pointed (e.g. $X$ is a manifold or $X$ admits a triangulation with $x$ as a vertex), then the evaluation at $x$, $\ev_x: \Map(X, Y) \to X$, is a fibration with fiber $\ev_x^{-1}\{y\} = \Map_*(X,Y)$.
\end{thm}

Fibrations $p: E \to B$ admit a path-lifting result analogous to Lemma \ref{PathLiftingLem}, but instead of the lift being unique, it is only unique up to homotopies preserving the starting point and the composition with $p$. Similarly to Definition \ref{MonodromyDef}, this gives rise to the \emph{monodromy action} of $\pi_1(B,b)$ on the path components of $p^{-1}\{b\}$. As the path components of $E$ are precisely the orbits of the monodromy action, we obtain the following result.

\begin{cor}\label{PointedVsUnpointedClassCor}
If $(X,x)$ and $(Y,y)$ are compactly generated pointed spaces and $(X,x)$ is well pointed, then there exists a natural isomorphism
$$[X, Y] \cong [X, Y]_* / \pi_1(Y,y).$$
\end{cor}

\begin{ex}
If $(S^n, e_0)$ is the $n$-sphere pointed at the north pole and $(X,x)$ is a compactly generated pointed space, then $[S^n, X]_* \cong \pi_n(X,x)$, and the monodromy action coincides with the usual $\pi_1$-action on the $n^{th}$ homotopy group.
\end{ex}

\section{Mathematical notation}

\begin{center}
	\begin{tabularx}{\columnwidth}{ p{3cm}  X }
    Symbol  & Meaning \\
    \hline
    $\cong$ & Isomorphism or a homeomorphism. \\
    $\simeq$ & Homotopy equivalence. \\
    $[x]$ & Equivalence class of $x$. \\
    $\{x\}$ & One-element set containing $x$. \\
    $X \to Y$ & Map from $X$ to $Y$. \\
    $X \hook Y$ & Injective map $X \to Y$. \\
    $x \mapsto f(x)$ & Formulaic description of a map $f: X \to Y$. Also known as an anonymous function. \\
    $f^{-1} S$ & Inverse image of $S \subset Y$ in a map $f : X \to Y$. \\
    $[X, Y]$ & Homotopy classes of continuous maps $X \to Y$. \\
    $[X, Y]_*$ & Basepoint-preserving homotopy classes of pointed maps $(X,x) \to (Y,y)$. The basepoints $x \in X$ and $y \in Y$ are usually omitted from notation. \\
    $\pi_n(X,x)$ & $n^{th}$ homotopy group of $X$ based at $x \in X$. Isomorphic as a set to $[S^n, X]_*$. \\
    $\Deck(\wtil X / X)$ & Group of deck transformations of a covering space $\wtil X \to X$. \\
    $X \coprod Y$ & Disjoint union. \\
    $X \vee Y$ & Wedge sum of pointed spaces $(X,x)$ and $(Y,y)$. \\
    $g.x$ & Given an action of a group $G$ on a space or a set $X$, this denotes the result of acting by $g \in G$ on $x \in X$. \\
    $X/G$ & Quotient space or set. Also called the set or the space of orbits. 
    \end{tabularx}
\end{center}

\bibliographystyle{apsrev4-1}
\bibliography{references}

\end{document}